\documentclass{IEEEtran}
\usepackage{amsmath,amstext,amsthm,amssymb}
\usepackage{latex8}
\usepackage{times}

\usepackage{amsmath,amstext,amssymb,epsf}
\usepackage{epsfig}
\usepackage{verbatim}
\usepackage{pslatex}

\newtheorem{coro}{\sc Corollary}
\newtheorem{nota}{\sc Notation}
\newtheorem{defin}{\sc Definition}
\newtheorem{rem}{\sc Remark}
\newtheorem{cla}{\sc Claim}
\newtheorem{ex}{\sc Example}
\newenvironment{remark}{\begin{rem}}{\hspace*{\fill}$\diamondsuit$\end{rem}}
\newenvironment{example}{\begin{ex}}{\hspace*{\fill}$\Diamond$\end{ex}}

\title{\Huge Compression-based Similarity}
\author{Paul M.B. Vit\'anyi\\CWI, Amsterdam, The Netherlands\\
({\em Invited Lecture})\thanks{Affiliation: National Research Center for Mathematics and Computer Science in the Netherlands (CWI). 
Address: CWI, Science Park 123, 1098XG Amsterdam,
The Netherlands. Email: Paul.Vitanyi@cwi.nl}}

\begin{document}
\maketitle

\begin{abstract}
First we consider pair-wise distances 
for literal objects consisting of finite binary
files. These files are taken to contain all of their
meaning, like genomes or books.
The distances are based on compression of the objects concerned,
normalized, and can be viewed as similarity distances.
Second, we consider pair-wise distances between
names of objects, like ``red'' or ``christianity.'' In this case the distances
are based on searches of the Internet. 
Such a search can be performed by any search  
engine that returns aggregate page counts. We can extract a code length 
from the numbers returned, use the same formula as before, 
and derive a similarity
or relative semantics between names for objects. The theory is based
on Kolmogorov complexity. We test both similarities
extensively experimentally.
\end{abstract}

\section{Introduction}
In pattern recognition, learning, and data mining one obtains
information from information-carrying objects. This involves
an objective definition of the information
in a single object, the information to go from one object to
another object in a pair of objects, the information to go from
one object to any other object in a multiple of objects,
and the shared information between objects.

The notion of
Kolmogorov complexity \cite{Ko65} is an objective measure
for the information in an
a {\em single} object, and information distance measures the information
between a {\em pair} of objects \cite{BGLVZ}. This leads to
the notion of similarity we shall explore below.

Objects can be given literally, like the literal
four-letter genome of a mouse,
or the literal text of {\em War and Peace} by Tolstoy. For
simplicity we take it that all meaning of the object
is represented by the literal object itself. Objects can also be
given by name, like ``the four-letter genome of a mouse,''
or ``the text of {\em War and Peace} by Tolstoy.'' There are
also objects that cannot be given literally, but only by name
and acquire their meaning from their contexts in background common
knowledge in humankind, like ``home'' or ``red.''
In the literal setting, similarity of objects can be established
by feature analysis, one type of similarity per feature.
In the abstract ``name'' setting, all similarity must depend on 
background knowledge and common semantics relations,
which is inherently subjective and ``in the mind of the beholder.'' 

\begin{remark}
\rm
As an aside, in many applications we are interested in
shared information between {\em many} objects
instead of just a pair of objects. For example, in customer reviews of
gadgets, in blogs about public happenings,
in newspaper articles about the same occurrence, we are interested
in the most comprehensive one or the most specialized one.
Thus, we want to extend the information distance
measure from pairs to multiples. This approach was introduced in \cite{Li08}
while most of the theory is developed in \cite{Vi11}.
\end{remark}

\section{Similarity}
All data are created equal but some data are more alike than others.
We have proposed methods expressing this alikeness,
using a new similarity metric based on compression.
It is parameter-free in that it
doesn't use any features or background knowledge about the data, and can without
changes be applied to different areas and across area boundaries.
It is universal in that it approximates the parameter
expressing similarity of the dominant feature in all pairwise
comparisons.
It is robust in the sense that its success appears independent
from the type of compressor used (among equally good compressors).
The clustering we use is hierarchical clustering in dendrograms
based on a new fast heuristic for the quartet method \cite{CV11}.
If we consider $n$ objects, then we find $n^2$ pairwise distances.
These distances are between natural data. We let the data decide for
themselves, and construct a hierarchical clustering of the $n$ objects
concerned. For details see the cited reference.
The method takes the  $n \times n$ distance matrix as input, and
yields a dendrogram with the $n$ objects as leaves (so the
dendrogram contains $n$ external nodes or leaves and $n-2$ internal nodes.
We assume $n \geq 4$.
The method is available as an open-source software tool, \cite{Ci03}.

\subsection{Feature-Based Similarities}
We are presented with unknown data and
the question is to determine the similarities among them
and group like with like together. Commonly, the data are
of a certain type: music files, transaction records of ATM machines,
credit card applications, genomic data. In these data there are
hidden relations that we would like to get out in the open.
For example, from genomic data one can extract
letter- or block frequencies (the blocks are over the four-letter alphabet);
 from music files one can extract
various specific numerical features,
related to pitch, rhythm, harmony etc.
One can extract such features using for instance
Fourier transforms~\cite{TC02} or wavelet transforms~\cite{GKC02},
to quantify parameters expressing similarity.
The resulting vectors corresponding to the various files are then
classified or clustered using existing classification software, based on
various standard statistical pattern recognition classifiers~\cite{TC02},
Bayesian classifiers~\cite{DTW97},
hidden Markov models~\cite{CV01},
ensembles of nearest-neighbor classifiers~\cite{GKC02}
or neural networks~\cite{DTW97,Sc01}.
For example, in music one feature would be to look for rhythm in the sense
of beats per minute. One can make a histogram where each histogram
bin corresponds to a particular tempo in beats-per-minute and
the associated peak shows how frequent and strong that
particular periodicity was over the entire piece. In \cite{TC02}
we see a gradual change from a few high peaks to many low and spread-out
ones going from hip-hip, rock, jazz, to classical. One can use this
similarity type to try to cluster pieces in these categories.
However, such a method requires specific and detailed knowledge of
the problem area, since one needs to know what features to look for.

\subsection{Non-feature Similarities}
Our aim
is to capture, in a single similarity metric,
{\em every effective distance}:
effective versions of Hamming distance, Euclidean distance,
edit distances, alignment distance, Lempel-Ziv distance,
and so on.
This metric should be so general that it works in every
domain: music, text, literature, programs, genomes, executables,
natural language determination,
equally and simultaneously.
It would be able to simultaneously detect {\em all\/}
similarities between pieces that other effective distances can detect
seperately.

Such a ``universal'' metric
was co-developed by us as a normalized
version of the ``information metric'' of \cite{LV08,BGLVZ}.
There it was shown that the information metric minorizes up to a constant all
effective distances satisfying a mild density requirement (excluding
for example distances that are 1 for every pair $x,y$ such that $x \neq y$).
This justifies the notion that the information distance is universal.

We may be interested what happens
in terms of properties or features of the 
pair of objects analyzed, say $x$ and $y$.
It can be shown that the information distance captures every property 
of which the Kolmogorov complexity
is logarithmic in the length of $\min\{|x|,|y|\}$. If those lengths go to
infinity, then logarithm of those lengths go to infinity too.
In this case the information distance captures every property. 

This information distance (actually a metric)
is normalized so that the resulting
distances are in $[0,1]$ and can be shown to retain the metric
property, \cite{Li02}. (A nonoptimal precursor
was given in \cite{LBCKKZ01}.) The result is the 
``normalized information distance'' (actually a metric).
All this is in terms of Kolmogorov complexity \cite{LV08}.
 
Intuitively, two objects are deemed close if
we can significantly ``compress'' one given the information
in the other, the intuition being that if two pieces are more similar,
then we can more succinctly describe one given the other.
The normalized information distance
 discovers all effective similarities in the sense that if two
objects are close according to some effective similarity, then 
they are also close according to the normalized information distance.

Put differently, the normalized information distance represents
similarity according to the dominating shared feature between
the two objects being compared.
In comparisons of more than two objects, 
different pairs may have different dominating features.
For every two objects,
this normalized information metric distance zooms in on the dominant
similarity between those two objects
 out of a wide class of admissible similarity
features. 
Since the normalized information distance also satisfies the metric 
(in)equalities, and takes values in $[0,1]$,
it may be called {\em ``the'' similarity metric}.

\subsection{Normalized Compression Distance}
Unfortunately, the universality of the normalized information distance
comes at the price of noncomputability. In fact, the normalized 
information distance is not even semicomputable (this is weaker than
computable) and there is no semicomputable function at a computable
distance of it \cite{TTV11}.
But since the Kolmogorov
complexity of a string or file is the length 
of the ultimate compressed version of that
file, 
we can use real data compression programs to approximate the Kolmogorov
complexity.
Therefore, to apply this ideal precise mathematical theory in real life,
we have to replace the use of  the noncomputable
Kolmogorov complexity by an approximation
using a standard real-world compressor. 
Starting from the normalized information distance, 
if $Z$ is a compressor and we use $Z(x)$
to denote the length of the compressed version of a string $x$,
then we arrive at the {\em Normalized Compression Distance}:
\begin{equation}\label{eq.ncd}
 NCD(x,y) = \frac{Z(xy) - \min(Z(x),Z(y))}{\max(Z(x),Z(y))},
\end{equation}
where for convenience we have replaced the pair $(x,y)$ in the formula
by the concatenation $xy$, and we ignore logarithmic terms
in the numerator and denominator,
see \cite{Li02,CV04}.
In \cite{CV04} we propose axioms to capture the real-world setting,
and show that \eqref{eq.ncd}
approximates optimality.
Actually, the
NCD is a family of compression functions parameterized 
by the given data
compressor $Z$. 
\begin{example}\label{example.genome}
{\sc (Phylogeny)}\index{phylogeny}
\rm
One cannot find
more appropriate data than DNA sequences to test our theory. 
A DNA sequence is a finite string over a 
4-letter alphabet $\{A,C,G,T \}$. We used the entire
mitochondrial genomes of 20 mammals, each of about 18,000 base pairs,
to test a hypothesis about the  Eutherian orders.
It has been hotly debated in biology which two of the 
three main placental mammalian groups,
primates, ferungulates, and rodents, are more closely related.
One cause of the debate is that in the analysis of the genomics
the standard maximum likelihood
method, which depends on the multiple alignment of sequences corresponding to
an individual protein, gives (rodents, (ferungulates, primates)) for half of
method, which depends on the multiple alignment of sequences corresponding to
an individual protein, gives (rodents, (ferungulates, primates)) for half of
the proteins in the mitochondrial genome, and 
(ferungulates, (primates, rodents)) for the other half.

\begin{figure}[ht]
\begin{center}
\hfill\ \psfig{figure=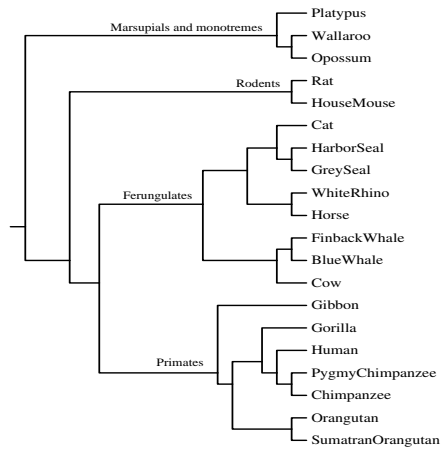,width=2.8in,height=2.9in} \hfill\
\caption{The evolutionary tree built from complete mammalian mtDNA
sequences}
\label{tree-mammal}
\end{center}
\end{figure}         

In recent years, as a result of more sophisticated methods, 
together with biological
evidence, it is believed that (rodents, (ferungulates, primates))
reflects the true evolutionary history. We confirm this from
the whole-genome perspective using the NCD distance.
We use the complete mitochondrial genome sequences 
from the following 20 species:
rat ({\em Rattus
norvegicus}), house mouse ({\em Mus musculus}), gray (or grey) 
seal ({\em Halichoerus grypus}), harbor seal ({\em Phoca vitulina}), cat ({\em
Felis catus}), white rhino ({\em Ceratotherium simum}), horse ({\em
Equus caballus}), finback whale ({\em Balaenoptera physalus}), blue
whale ({\em Balaenoptera musculus}), cow ({\em Bos taurus}), gibbon
({\em Hylobates lar}), gorilla ({\em Gorilla gorilla}), human ({\em
Homo sapiens}), chimpanzee ({\em Pan troglodytes}), pygmy chimpanzee
({\em Pan paniscus}), orang\-utan ({\em Pongo pygmaeus}), Sumatran
orangutan ({\em Pongo pygmaeus abelii}), with opossum ({\em Didelphis
virginiana}), wallaroo ({\em Macropus robustus}), and platypus ({\em
Ornithorhynchus anatinus}) as the outgroup. 

For every pair of mitochondrial genome sequences
$x$ and $y$, evaluate the formula in Equation~\ref{eq.ncd}
using a special-purpose DNA sequence compressor DNACompress,
or a good general-purpose compressor such as PPMZ.
The resulting distances are the entries in an
$n \times n$ distance matrix. Constructing a phylogeny tree
from the distance matrix, using common
tree-reconstruction software, gives the tree in Figure~\ref{tree-mammal}.
This tree confirms the accepted hypothesis of 
(rodents, (primates, ferungulates)), and every single branch of the 
tree agrees with the current biological classification.
\end{example}

\begin{example}
{\sc (Hierarchical clustering)}\index{clustering!hierarchical}
\rm
The normalized compression distance has been used to fully automatically
reconstruct language and phylogenetic trees as above. It can, and has, also be used
for a plethora of new applications of general clustering and classification of natural data
in arbitrary domains, for clustering of heterogeneous data, and for
anomaly detection across domains. It has further been applied to authorship attribution,
stemmatology, music classification, Internet knowledge discovery,
to analyze network traffic and cluster computer worms and viruses,
software metrics and obfuscation, web page authorship, topic and
domain identification, hurricane risk assessment, ortholog detection,
and clustering fetal heart rate tracings.
We test gross classification of files
based on heterogeneous data of markedly different file types:
(i) four mitochondrial gene sequences, from a black bear, polar bear,
fox, and rat obtained from the GenBank Database on the worldwide web;
(ii) four excerpts from the novel { \em The Zeppelin's Passenger} by
E.~Phillips Oppenheim, obtained from the Project Gutenberg Edition
on the worldwide web;
(iii) four MIDI files without further processing, two works by Jimi Hendrix and
two movements from Debussy's ``Suite Bergamasque,'' downloaded from various
repositories on the
worldwide web;
(iv) two Linux x86 ELF executables (the {\em cp} and {\em rm} commands),
copied directly from the RedHat 9.0 Linux distribution; and
(v)  two compiled Java class files, generated directly.
The program correctly classifies each of the different types
of files together with like near like. The result is reported
in Figure~\ref{figfiletypes}.
This experiment shows the power and universality of the method:
no features of any specific domain of application are used.
We believe that there is no other method known that can cluster
data that are so heterogeneous this reliably.
\begin{figure}[htb]
\begin{center}
\epsfysize=4in
\leftline{\hskip0.5pc\epsfbox{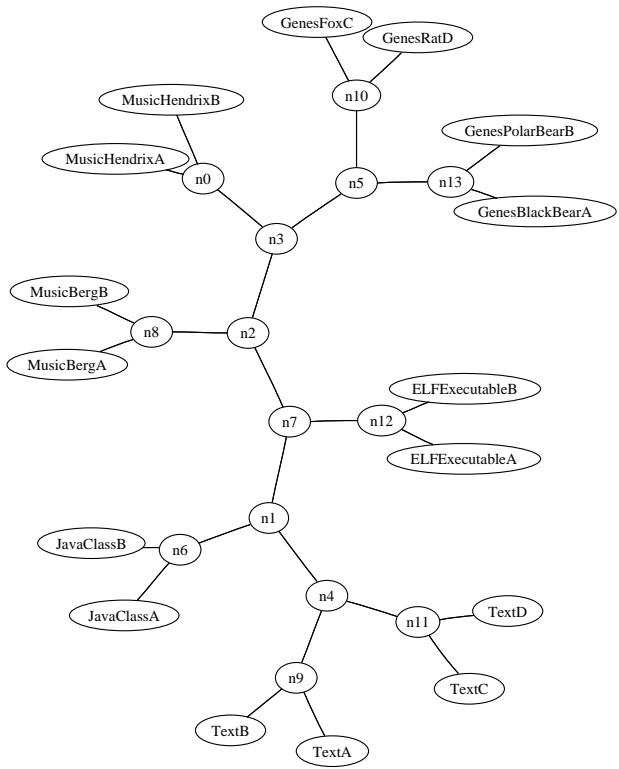}}
\end{center}
\caption{Clustering of heterogeneous file types}\label{figfiletypes}
\end{figure}
Researchers from the data-mining community noticed that this methodology
is in fact a parameter-free, feature-free, data-mining tool.
They have experimentally tested a closely related metric on a large variety
of sequence benchmarks. Comparing the compression-based method with 51 major
parameter-loaded  methods
found in the 7 major data-mining conferences
(sigkdd, sigmod, icdm, icde, ssdb, vldb,
pkdd, and pakdd) in 1994--2004,
on every database of time sequences used,
ranging from heartbeat signals to stock market curves,
they established clear superiority of the compression-based method for
clustering heterogeneous data, for anomaly detection,
and competitiveness in clustering domain data \cite{Ke04}.
\end{example}

\subsection{NCD in Retrospect} In \cite{CV04} it is proved that the 
NCD is a metric.
The compression-based NCD method 
establishes a similarity metric \eqref{eq.ncd} among objects
given as finite binary strings. 
It has been applied to
objects like genomes, music pieces in MIDI format, computer programs
in Ruby or C, pictures in simple bitmap formats, or time sequences such as
heart rhythm data, heterogenous data and anomaly detection.
This method is feature-free in the sense
that it doesn't analyze the files looking for particular
features; rather it analyzes all features simultaneously
and determines the similarity between every pair of objects
according to the most dominant shared feature. 

The crucial
point is that the method analyzes the objects themselves.
This precludes comparison of abstract notions or other objects
that don't lend themselves to direct analysis, like
emotions, colors, Socrates, Plato, Mike Bonanno and Albert Einstein.

\section{Web-based Similarity}
To make computers more intelligent one would like
to represent meaning in computer-digestable form.
Long-term and labor-intensive efforts like
the {\em Cyc} project \cite{Le95} and the {\em WordNet}
project \cite{Miea} try to establish semantic relations
between common objects, or, more precisely, {\em names} for those
objects. The idea is to create
a semantic web of such vast proportions that rudimentary intelligence
and knowledge about the real world spontaneously emerges.
This comes at the great cost of designing structures capable
of manipulating knowledge, and entering high
quality contents in these structures
by knowledgeable human experts. While the efforts are long-running
and large scale, the overall information entered is minute compared
to what is available on the world-wide-web.

The rise of the world-wide-web has enticed millions of users
to type in trillions of characters to create billions of web pages of
on average low quality contents. The sheer mass of the information
available about almost every conceivable topic makes it likely
that extremes will cancel and the majority or average is meaningful
in a low-quality approximate sense. Below, we give a general
method to tap the amorphous low-grade knowledge available for free
on the world-wide-web, typed in by local users aiming at personal
gratification of diverse objectives, and yet globally achieving
what is effectively the largest semantic electronic database in the world.
Moreover, this database is available for all by using any search engine
that can return aggregate page-count estimates like Google for a large
range of search-queries.

While the previous NCD method that compares the objects themselves using
\eqref{eq.ncd} is
particularly suited to obtain knowledge about the similarity of
objects themselves, irrespective of common beliefs about such
similarities, we now develop a method that uses only the name
of an object and obtains knowledge about the similarity of objects
by tapping available information generated by multitudes of
web users.
The new method is useful to extract knowledge from a given corpus of
knowledge, in this case the world-wide-web accessed by a search engine
returning aggregate page counts, but not to
obtain true facts that are not common knowledge in that database.
For example, common viewpoints on the creation myths in different
religions
may be extracted by the web-based method, but contentious questions
of fact concerning the phylogeny of species can be better approached
by using the genomes of these species, rather than by opinion.
The approach below was proposed by \cite{CV07}.

\subsection{The Search Distribution}
\label{sect.google}
Let the set of singleton {\em search terms}
be denoted by ${\cal S}$. In the sequel we use both singleton
search terms and doubleton search terms $\{\{x,y\}: x,y \in {\cal S} \}$.
Let the set of web pages indexed (possible of being returned)
by search engine be $\Omega$. The cardinality of $\Omega$ is denoted
by $M=|\Omega|$. 
Assume that a priori all web pages are equiprobable, with the probability
of being returned  being $1/M$.  A subset of $\Omega$
is called an {\em event}. Every {\em  search term} $x$ 
defines a {\em singleton event} ${\bf x} \subseteq \Omega$ of web pages
that contain an occurrence of $x$ and are returned by the search engine
if we do a search for $x$.
Let $L: \Omega \rightarrow [0,1]$ be the uniform mass probability
function.
The probability of
such an event ${\bf x}$ is $L({\bf x})=|{\bf x}|/M$.
 Similarly, the {\em doubleton event} ${\bf x} \bigcap {\bf y}
\subseteq \Omega$ is the set of web pages returned 
if we do a search for pages containing both search term $x$ and
search term $y$.
The probability of this event is $L({\bf x} \bigcap {\bf y})
= |{\bf x} \bigcap {\bf y}|/M$.
We can also define the other Boolean combinations: $\neg {\bf x}=
\Omega \backslash {\bf x}$ and ${\bf x} \bigcup {\bf y} =
\Omega \backslash ( \neg {\bf x} \bigcap \neg {\bf y})$, each such event
having a probability equal to its cardinality divided by $M$.
If ${\bf e}$ is an event obtained from the basic events ${\bf x}, {\bf y},
\ldots$, corresponding to basic search terms $x,y, \ldots$,
by finitely many applications of the Boolean operations,
then the probability $L({\bf e}) = |{\bf e}|/M$.

These events capture in a particular sense
all background knowledge about the search terms concerned available
(to search engine) on the web. Therefore, it is natural
to consider code words for those events
as coding this background knowledge. However,
we cannot use the probability of the events directly to determine
a prefix code such as the Shannon-Fano code \cite{LV08}.
The reason is that
the events overlap and hence the summed probability exceeds 1.
By the Kraft inequality \cite{LV08} this prevents a
corresponding Shannon-Fano code.
The solution is to normalize:
We use the probability of the events to define a probability
mass function over the set $\{\{x,y\}: x,y \in {\cal S}\}$
of  search terms, both singleton and doubleton.
Define
\[
 N= \sum_{\{x,y\} \subseteq {\cal S}} |{\bf x} \bigcap
{\bf y}|,
\]
counting each singleton set and each doubleton set (by definition
unordered) once in the summation.
Since every web page that is indexed by contains at least
one occurrence of a search term, we have $N \geq M$. On the other hand,
web pages contain on average not more than a certain constant $\alpha$
search terms. Therefore, $N \leq \alpha M$.
Define
\begin{align}\label{eq.gpmf}
&g(x) = L({\bf x}) M/N =|{\bf x}|/N
\\&
\nonumber
g(x,y) =  L({\bf x} \bigcap {\bf y}) M/N =|{\bf x} \bigcap {\bf y}|/N.
\end{align}
Then, $\sum_{x \in {\cal S}} g(x)+ \sum_{x,y \in {\cal S}} g(x,y) = 1$.
Note that $g(x,y)$ is not a conventional joint distribution
since possibly $g(x) \neq \sum_{y \in {\cal S}} g(x,y)$.
Rather, we consider $g$ to be a probability mass
function over the sample space $\{ \{x,y\}: x,y \in {\cal S} \}$.
This $g$-distribution changes over time,
and between different samplings
from the distribution. But let us imagine that $g$ holds
in the sense of an instantaneous snapshot. The real situation
will be an approximation of this.
Given the search machinery, these are absolute probabilities
which allow us to define the associated Shannon-Fano code for
both the singletons and the doubletons.

\subsection{Normalized Web Distance}
The {\em web code} length $G$
is defined by
\begin{align}\label{eq.gcc}
&G(x)= \log 1/g(x)
\\&
\nonumber
G(x,y)= \log 1/g(x,y) .
\end{align}
In contrast to strings $x$ where the complexity $Z(x)$ represents
the length of the compressed version of $x$ using compressor $Z$, for a search
term $x$ (just the name for an object rather than the object itself),
the  code of length $G(x)$ represents the shortest expected
prefix-code word length of the associated event ${\bf x}$.
The expectation
is taken over the distribution $p$.
In this sense we can use the distribution as a compressor
for ``meaning'' associated with the search terms.
The associated 
{\em normalized web distance} (NWD) is defined
just as \eqref{eq.ncd} with the search engine in the role of compressor
yielding code lengths $G(x), G(y)$
for the singleton search terms $x,y$ being compaired
and a code length $G(x,y)$ for the doubleton pair $(x,y)$, by
\begin{equation}\label{eq.NGD}
 NWD(x,y)=\frac{G(x,y) - \min(G(x),G(y))}{\max(G(x),G(y))}.
\end{equation}
This $NWD$ 
uses the background knowledge on the web as viewed by the search engine
as conditional information.

\subsection{Searching the Web for Knowledge}
Many web search engines index more than
twenty-five billion pages on the web. Each such page can be
viewed as a set of index terms. A search for a particular index term 
(in 2004)
say ``horse'', returned a certain number of hits (web pages where
this term occurred), say 46,700,000. The number of hits for the
search term ``rider'' is, say, 12,200,000. It is also possible to search
for the pages where both ``horse'' and ``rider'' occur. This gives,
say, 2,630,000 hits.
The search engine searched at that time, say, 8,058,044,651 web pages.
The 
same formula as \eqref{eq.NGD} can be written
in terms of frequencies as
\begin{equation}\label{eq.ngd}
NWD(x,y) = \frac{  \max \{\log f(x), \log f(y)\}  - \log f(x,y) \}}{
\log N - \min\{\log f(x), \log f(y) \}},
\end{equation}
and if $f(x),f(y)>0$ and $f(x,y)=0$ then $NWD(x,y)= \infty$.
It is easy to see that
\begin{enumerate}
\item
$NWD(x,y)$ is undefined for  $f(x)=f(y)=0$;
\item
$NWD(x,y) = \infty$ for $f(x,y)=0$ and either or both $f(x)>0$
and $f(y)>0$; and
\item
$ NWD(x,y) \geq 0$ otherwise.
\end{enumerate}
Our experimental results suggest that every reasonable
(greater than any $f(x)$) value can be used for the normalizing factor  $N$,
and our
results seem  in general insensitive to this choice.  In our software, this
parameter $N$ can be adjusted as appropriate, and we often use $M$ for $N$.
In the \cite{CV07} we analyze the mathematical properties of NWD,
and  prove the universality of the search engine distribution.
We show that the NWD is not a metric, in contrast to the NCD.
The generic example showing the nonmetricity of semantics (and therefore
the NWD) 
is that a man is close to a centaur,
and a centaur is close to a horse, but a man is very different from
a horse.

With the hit numbers above, using $M$ for $N$,
we can compute from \eqref{eq.NGD} that
\[
NWD(horse,rider)
\approx 0.443.
\]
The NWD formula itself \eqref{eq.ngd} is {\em scale-invariant} 
in the sense that
if $M$ doubles and so do
the $f$-frequencies then the result stays the same. 
\begin{example}
{\sc (Classification)}
\rm
In cases in which the set of objects can be large,
in the millions, clustering cannot do us much good.
We may also want to do definite classification, rather than
the more fuzzy clustering.
One can use the NCD/NWD distances
as an oblivious feature-extraction technique to convert
generic objects into finite-dimensional vectors. 

(We have used this
technique to train a support vector machine
(SVM) based OCR system to classify handwritten digits
by extracting 80 distinct, ordered NCD features from each input image
in the manner explained below in the context of the NWD experiments.
For details about the SVM see \cite{Bu97}.
We achieved a handwritten single decimal digit recognition accuracy of 87\%.
The current state of the art for this problem,
after half a century of interactive feature-driven
classification research, is in the upper ninety percent level.
These experiments were benchmarked on the
standard  NIST Special Data Base 19.)

For classification using the NWD
distance, 
the setting is, say, a binary classification problem on examples represented
by search terms.  In this experiment,
we require a human expert to provide a list of
at least 40 {\em training words},
consisting of at least 20 positive examples and 20 negative examples,
to illustrate the
contemplated concept class.  The expert also provides, say,
six {\em anchor words} $a_1, \ldots , a_6$,
of which half are in some way related to the concept
under consideration.  Then, we use the anchor words to convert each
of the 40 training words $w_1 , \dots , w_{40}$
to 6-dimensional {\em training vectors} $\bar{v}_1 , \ldots , \bar{v}_{40}$.
The entry $v_{j,i}$ of $\bar{v}_j = (v_{j,1}, \ldots , v_{j,6})$
is defined as $v_{j,i} = NWD(w_j, a_i)$ ($1 \leq j \leq 40$,
$1 \leq i \leq 6$).
The training vectors are
then used to train an SVM to learn the concept. The test words
are classified using the same anchors and trained SVM model.
The LIBSVM software was used for all SVM experiments~\cite{CL01}.

In an experiment to learn prime numbers, we used the literal search terms
below (digital numbers and alphabetical words) in the Google search engine.\\
{\em Positive training examples}:
11, 13, 17, 19, 2,
23, 29, 3, 31, 37,
41, 43, 47, 5, 53,
59, 61, 67, 7, 71,
73. \\
{\em Negative training examples}:
10, 12, 14, 15, 16,
18, 20, 21, 22, 24,
25, 26, 27, 28, 30,
32, 33, 34, 4, 6,
8, 9.\\
{\em Anchor words}:
composite, number, orange, prime, record.\\
{\em Unseen test examples}:
The numbers 101, 103,
107, 109,
79, 83,
89,
97 were correctly classified as primes.
The numbers 36, 38,
40, 42,
44, 45,
46, 48,
49 were correctly classified as nonprimes.
The numbers 91 and 110 were false positives,
since they were incorrectly classified as primes.
There were no false negatives.
The accuracy on the test set is  $17/19 = 89.47\%$.
Thus, the method automatically learns to distinguish prime numbers from
nonprime numbers by example, using a search engine that knows nothing about
mathematics. Note that in this example we have to keep the numbers under,
say, 200, since larger numbers do not necessarily occur on the web in
the required multitude.

The NWD has been used for clustering, classification,
and translation of small samples \cite{CV07}. That reference
reports a massive experiment
comparing the performance of the NWD--SVM method with the human-expert-entered
information in the WordNet\index{WordNet} database \cite{Miea}.
They showed a mean accuracy of agreement of 87.25\%
of the NWD--SVM method with the WordNet semantic
concordance.
\end{example}

\section{Conclusion}
\label{sect.exp}
By now applications abound. 
See the many references to the papers
\cite{Li02,CV04,CV07} in Google Scholar. 
Below we treat some
favorite applications we are fimiliar with out of this multitude.

{\sc Applications of NCD:}
We and others performed
experiments in vastly different
application fields to test the quality and universality of the method.
The success of the method as reported below depends strongly on the
judicious use of encoding of the objects compared. Here one should
use common sense on what a real world compressor can do. There are
situations where our approach fails if applied in a
straightforward way.
For example: comparing text files by the same authors
in different encodings (say, Unicode and 8-bit version) is bound to fail.
For the ideal similarity metric  based on
Kolmogorov complexity as defined in \cite{Li02}
this does not matter at all, but for
practical compressors used in the experiments it will be fatal.
Similarly, in the music experiments we use symbolic MIDI
music file  format rather than wave format music files. The reason is that
the strings resulting from straightforward
discretizing the wave form files may be too sensitive to how we discretize.
Further research may overcome this problem.

The method is implemented and available as public software \cite{Ci03}.
This approach gives
the first completely automatic construction 
of the phylogeny tree based on whole mitochondrial genomes (above),
and a completely automatic construction of a language tree for over 50
Euro-Asian languages \cite{Li02},
detects plagiarism in student programming assignments
\cite{CFLMS04}, gives phylogeny of chain letters \cite{BLM03}, and clusters
music \cite{CWV04}.
Moreover, the method turns out to be more-or-less 
robust under change of the underlying
compressor-types: statistical (PPMZ), Lempel-Ziv based  dictionary (gzip),
block based (bzip2), or special purpose (Gencompress). Obviously the window
size matters, as well as how good the compressor is. For example, PPMZ
gives for mtDNA of the investigated species diagonal elements ($NCD(x,x)$)
between 0.002 and 0.006. The compressor bzip2 does considerably worse,
and gzip gives something in between 0.5 and 1 on the diagonal elements.
Nonetheless, for texts like books gzip does fine in our experiments;
the window size is sufficient and we do not use the diagonal elements.
But for genomics gzip is no good.

In \cite{CV04}
we report evidence of successful application in areas as diverse as
genomics, virology, languages, literature, music, handwritten digits,
astronomy, and
combinations of objects from completely different
domains (see example above), using statistical, dictionary, and block sorting compressors.
In genomics we presented new evidence for major questions
in Mammalian evolution, based on whole-mitochondrial genomic
analysis: the Eutherian orders (we gave this example above)
 and the Marsupionta hypothesis
against the Theria hypothesis.
Apart from the experiments reported in \cite{CV04}, the clustering by
compression method reported
in this paper  has recently been used in many different areas all over
the world. 
For example, how
to analyze network traffic and cluster computer worms and virusses~\cite{We04}.

{\sc Applications of NWD:}
 This method is  proposed in  \cite{CV07} to extract semantic
knowledge from the world-wide-web for both
supervised and unsupervised learning using a search engine
in an unconventional manner.  The approach is
novel in its unrestricted problem domain, simplicity of implementation,
and manifestly ontological underpinnings.  

Evidence is given of
elementary learning of the semantics of concepts, in
contrast to most prior approaches (outside of Knowledge Representation
research) that have neither the appearance nor the aim of dealing with ideas,
instead using abstract symbols that remain permanently ungrounded throughout
the machine learning application. 

The world-wide-web is the largest database on earth,
and it induces a
 probability mass function
 via page counts for search queries.
This distribution allows us to tap the latent semantic knowledge
on the web.
While in the NCD compression-based method
 one deals with the objects themselves,
in the NWD method we deal with just names for the objects.

In \cite{CV07}, as proof of principle, we demonstrate
positive correlations, evidencing an
underlying semantic structure, in both numerical symbol notations
and number-name words in a variety of natural languages
and contexts.
Next, we give applications in
(i) unsupervised hierarchical clustering, demonstrating the ability
to distinguish between colors and numbers, and
to distinguish between 17th century
Dutch painters;
(ii)
 supervised
concept-learning by example, using support vector machines,
demonstrating the ability to understand
electrical terms, religious terms,
emergency incidents, and by conducting
 a massive experiment in understanding
WordNet categories;
and (iii) matching of meaning, in an example of 
automatic English-Spanish translation.

An application that uses both the NCD and the NWD (and some derived
distance measures) is the competitive question-answer system in \cite{ZHZL08}.

\end{document}